# Deep Connection: Making Virtual Reality Artworks with Medical Scan Data

Marilène Oliver*
University of Alberta

Gary James Joynes†
 

Kumar Punithakumar
University of Alberta

Peter Seres
University of Alberta

**Abstract**

*Deep Connection* is an installation and virtual reality (VR) artwork made using full body 3D and 4D magnetic resonance (MR) scan datasets. When the user enters *Deep Connection*, they see a scanned body lying prone in mid-air. The user can walk around the body and inspect it, lie underneath and walk through it. The user can dive inside and see its inner workings, its lungs, spine, brain. The user can take hold of the figure's outstretched hand: holding the hand triggers the 4D dataset, making the heart beat and lungs breathe. When the user lets go the hand, the heart stops beating and the lungs stop breathing. *Deep Connection* creates a scenario where an embodied human becomes the companion for a virtual body. This paper maps the conceptual and theoretical framework for *Deep Connection* such as virtual intimacy and digitally mediated companionship. It also reflects on working with scanned bodies more generally in virtual reality by discussing transparency, the cyberbody versus the data body, as well as data privacy and data ethics. The paper also explains the technical and procedural aspects of the *Deep Connection* project with respect to acquiring scan data for the creation of virtual reality artworks.

## 1 Introduction

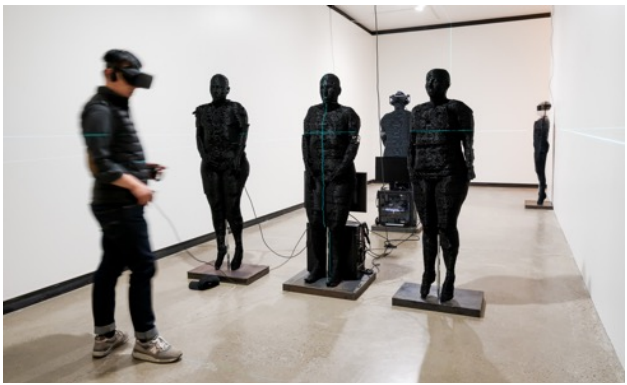

Figure 1: *Deep Connection* installation, lasercut coroplast, steel VR hardware, lasers, 2.5, x 1.8 x .5m, 2019.

*marilene@ualberta.ca
†info@clinker.ca

He bent down, put his head her stomach, shuffled though the spine, emerged from her brain and gasped "I couldn't breathe in there." The visibly affected man had just spent over ten minutes in the virtual reality (VR) artwork *Deep Connection*, made using a full body magnetic resonance (MR) scan dataset of Marilène Oliver's body. In *Deep Connection*, Oliver's virtual, translucent body floats prone in mid-air. It is possible to walk around and through Oliver's scanned body, to lean down and put your head inside it and see its inner forms and textures. Holding the hand makes the heart beat and lungs breathe.

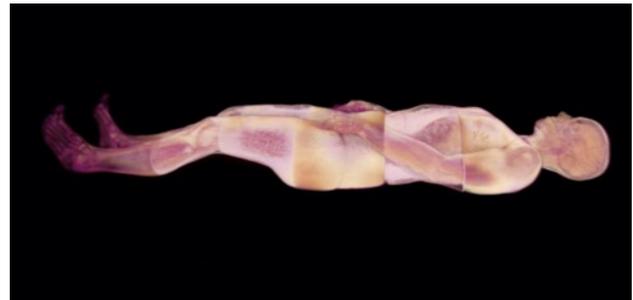

Figure 2: *Deep Connection* VR artwork as it appears when the user first puts on the headset.

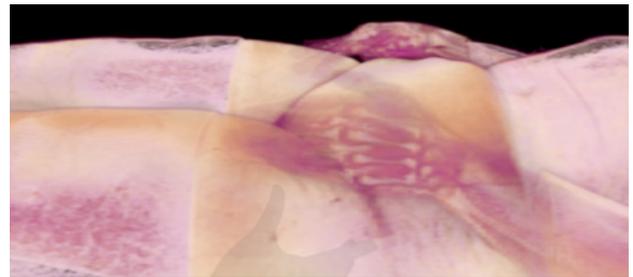

Figure 3: screen capture of *Deep Connection* showing user's hand reaching to hold the virtual hand.

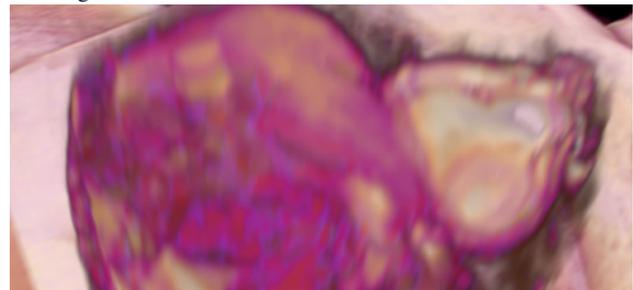

Figure 4: screen capture of *Deep Connection* showing heart and lungs.

The VR artwork is presented as part of an installation that includes sculptures made from the same scan data. VR hardware is embedded into these sculptures in the hope of invoking Heidegger's concept of a 'standing reserve,' where humans are destined to be both a *resource for* and *enablers of* technology [1]. When encountering conventional sculptural objects or figures, viewers generally stand back, give up space, and refrain from touching the works in question. Yet VR artworks demand the opposite. Viewers must participate; they put on a head set, move directly into the work, and interact with it, often by means of touch as well as vision (albeit mediated by VR headsets and controllers).

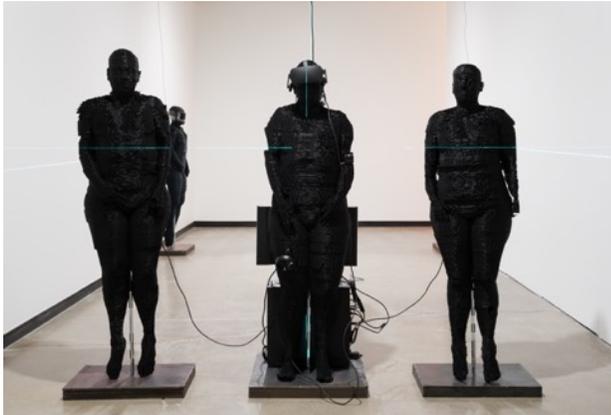

Figure 5. *Deep Connection* installation, lasercut coroplast, steel, VR hardware, lasers, 2.5, x 1.8 x .5m, 2019.

The title of the artwork *Deep Connection,* guides the viewer to consider the true motivation for making the work. In 2017, Oliver spent two weeks at her mother's bedside, holding her hand, watching her sleep, and observing the rise and fall of her struggled breath until she died. In recent years there has been a huge growth in the healthcare industry of automated care and tele-medicine [2][3]. The image of Oliver's mother being with a companion robot as she died rather than human touch was thus a primary motivator for making *Deep Connection*. *Deep Connection* inverts the scenario of an automation being a companion for human by making a human the companion for a virtual body. As long as the human continues to hold the hand of the virtual body, the 4D dataset is triggered so that the heart beats and the lungs breathe. If they let go, the heart stops beating, the lungs stop breathing.

Although made a year before the start of the COVID-19 pandemic, *Deep Connection* is also a sharp reminder of the reality that so many people have died and are still dying alone, isolated from their loved ones due to COVID-19 health protocols. During the pandemic we have heard and read numerous accounts of how nurses and palliative care workers have been holding iPads wrapped in plastic bags for patients on their deathbeds, facilitating Zoom calls so loved ones can say their last good bye [4]. In an article for *The Atlantic*, [5] Susan Zhang interviewed a number of people who facilitated end of life Zoom calls for patients and their families. The article includes several testimonies describing how spouses, siblings and children had to watch their closest relatives dying via a digitally mediated teleconferencing app. One of the interviewees, Stephanie Brook Kiser, explained however that some of the families preferred not to do this, and instead was asked her to hold the hand of the dying patient, they told her: "*we just want you to go to his bedside and hold his hand so that he knows he's not alone*."

## 2. 1 Radiology and Art

The first ever x-ray was also of the hand of a loved one. It was the left hand of Anne Bertha Röntgen, the wife of German engineer and physicist, Wilhem Röntgen who invented the x-ray. Since this first tender x-ray in 1895, a multitude of non-invasive technologies have been developed to render the body transparent and generate 'digital copies' of human bodies without having to cut them open. During the Renaissance, anatomists relied on artists to show the world what they had discovered [6]. In the nineteenth century, the subjective interpretation of anatomical subjects fell out of favour and "scientists hoped to eliminate artistic contamination" [7]. MR facilitates just that: the artist is no longer needed to render the interior body visible; digitized machines now do it in their place with apparent objectivity. Machines now "render the body, or more precisely the appearance of the body, into digital information, decomposing the body's fleshy complexity into the simple on/off logic of binary code" [8].Where anatomical art once told us to "Know Thyself" as mortal, or as a divine creation of God [9][10], the digital scan dataset suggests we "Know Thyself" as a virtual product [11], which, with VR, we can become totally immersed in and explore in heretofore unimaginable detail.

There are a number of artists who work with medical scans as part of their visual language and who are actively shaping the aesthetics of medical imaging [12]. Annie Cattrell [13] and Jane Prophet [14] have both worked sensitively with scan data to create sculptures that invite us to consider the emotional weight of scan data as a memento mori, or as a relic to treasure. Other artists such as Elizabeth Jameson [15] and Darian Stahl [16] work with scans as a way to express the lived experience of illness. British film maker Victoria Mapplebeck has also worked with scan data as part of her 2019 VR film *Waiting Room* which is based on her experience of living with breast cancer.

For Oliver, however, the main interest when working with scan data is to challenge and interrogate the increasing digitisation and datafication of the human. MR and other computerised radiological imaging technologies such as computer tomography (CT) and ultrasound are in essence digitally mechanised processes that measure the substance of the body at a cellular level and convert flesh to pixel, flesh to voxel, and flesh to xyz coordinate, offering what we believe to be a potent metaphor to think, see and know with in the Digital Age.

## 2. 2 Previous artworks with MR data

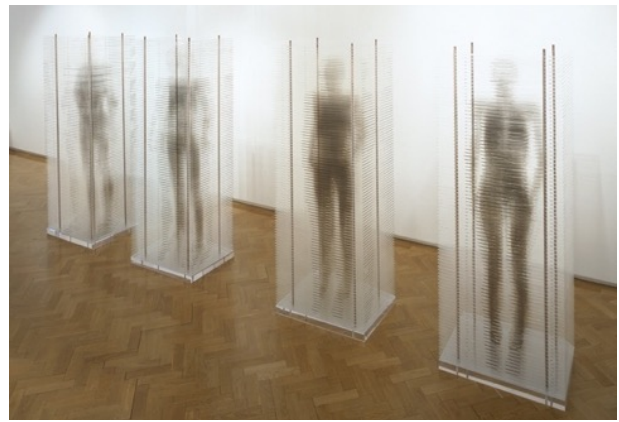

Figure 6. *Family Portrait* installation, screenprints in bronze ink on acrylic, each figure 50x70x190cm, 2003

Oliver has been working with medical imaging data such as MR and CT scans as an artist for almost 20 years. One of Oliver's earliest works was *Family Portrait,* a sculptural installation for which she arranged to have each of her family members MR scanned. Later Oliver screen printed the scans onto sheets of clear acrylic and stacked them to create a sculptural installation. Oliver was drawn to the MR scan then, as she is now, because it is one of the most intimate and precise ways of capturing and digitizing a human being. MR, CT and other scanning technologies resemble photographs, in that they are indexical, and they promise access to interiority by making our invisible interior visible.

When Oliver made *Family Portrait,* she was challenging post humanist notions such as those put forward by Hans Moravec in *Mind Children* [17]. In this book, Moravec speculates that in order for humans to survive in the Digital Age they will have to 'download their consciousness to the datascape.' Making *Family Portrait* was asserting the need to also consider how embodiment is digitised and challenging the mind/body dualism in Moravec's proposal. It also sentimental, exploring the desire to preserve loved ones - suggesting that scan data could be a future relic to cherish and covet.

In 2003 when the *Family Portrait* scans were acquired it took almost 90 minutes to scan a full body scan at 20mm intervals. For *Family Portrait* these 20 mm gaps between the slices were crucial in that they exposed the emptiness of the scanned body and the impossibility of digitising the human. Almost 20 years since *Family Portrait*, MR scanning technology has advanced so much that it is now possible to acquire much, much higher resolution datasets in far, far less time. Scan resolution is now so high that datasets can be volumetrically rendered in virtual reality to resemble 3D and 4D semi-transparent photographs. Transparency is key element to a large number of works Oliver has made using scan data as it allows the light to pass through the data and make it visible. The cross sections were printed onto sheets onto clear acrylic and the light that passed into and through the gaps permitted the illusion of a whole body.

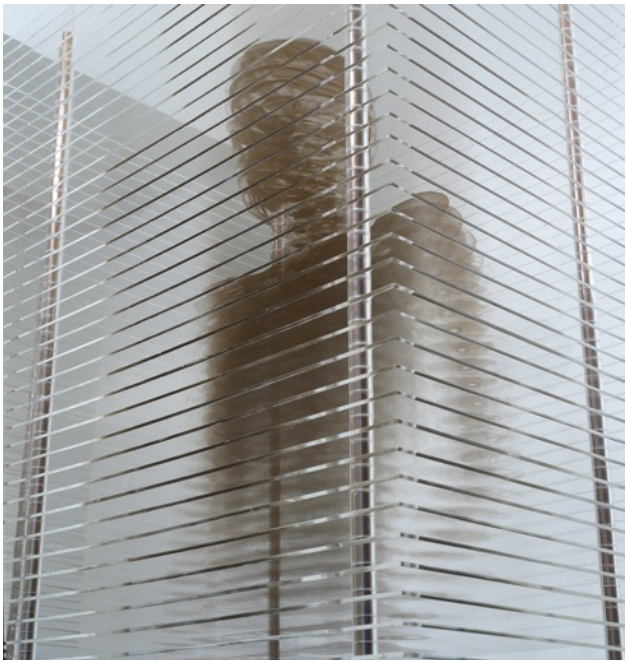

Figure 7. *Family Portrait* detail showing gaps between scans

For *Deep Connection*, transparency was also key to creating a successful data visualisation. Each voxel/sub-mm cube of data has a transparency level as well as a red, green and blue (RGB) value that allows the ray-casting technology to make the inside of the dataset visible. In *Deep Connection,* the fat layer has a high transparency value in order to expose the muscles and bones beneath it. Similarly, the transparency values make it possible for the viewer to see the heart beating and lungs breathing when they hold the hand. Or, to be more precise, the air in the lungs is completely transparent and what is visualised are the organs moving up and down. Working with transparency in *Deep Connection* has more than just a technical benefit however: it is also a key conceptual element of the project. Transparency is a major feature of the digital age – the more information/data we give away to the black box of technology, the more transparent we become [18][19].

**2.3 Transparency**
Lisa Cartwright, Jose Van Dijck, and Jenny Slatman have written extensively around the myths and implications of contemporary medical imaging and the transparent bodies they produce. From challenging the myth that rendering the inside of the body visible gives access to psychological interiority [20], to insisting that medical scanning methods be considered as techniques entangled in a history of mass media that promote problematic ideals of perfectibility and modifiability [7], to Cartwright's discussions of contemporary medical transparency as an extension of Foucault's concept of a disciplining and surveilling male medical gaze [21], this scholarship generally challenges the notion that medical transparency promises a greater poetic knowing of ourselves. In *The Four-Dimensional Human*, Laurence Scott [22] explores both how social media sites seduce us to carve into digital stone what we previously would only have whispered to a trusted confidante, and how sites such as Air B'n'B facilitate a new fish bowl transparency into our domestic lives, where the previously private and intimate space of our homes are now made visible (and purchasable) to all those with Internet access. In *The Circle*, David Eggers [23] also presents 'transparent' beings emerging as a result of social media: Mae, who becomes 'transparent' by allowing her life to be constantly broadcast by webcams in her home and carried on her body, as well as having the inside of her body constantly monitored and tracked by smart devices. A transparent shark whose every act of consumption and digestion are spectacularly exposed and celebrated in the foyer of the social media corporation in which Mae works.

Although Egger's *Circle* is fiction, it reflects fact. Biometric technologies have developed significantly in recent years and digitising the body with imaging technologies such as retinal scanning, fingerprint digitization in addition to face scanning so that they be used to gatekeep other data (such as social security, vaccination and banking data) is *big, big business*. Google for example have in recent years through its parent company, Alphabet, acquired 23andMe (DNA testing), Oscar Health (insurance), Doctor on Demand (virtual care), Verily (bio research), Calico (anti-aging research) and many other health related companies [24]. Google's consumer facing Nest, best known for its smart thermostats, is broadening its reach through the development of consumer health monitoring technologies following its purchase of Senosis Health in 2018 [25]. Google also acquired Fitbit in 2019 [26]. Google's *Care Studio* which launched as a pilot in February 2021, is a tool for clinicians that harmonizes data from disparate electronic health records into a single dashboard [27] that would be 'unlocked' by a patient's retinal scan. I doubt even post-humanist thinkers such as Moravec could have dreamed this terrifying level of data acquisition and data aggregation, especially not from a for-profit mega corporation such as Google. More alarmingly, governments are also deploying these technologies: China's social credit system aggregates data from shopping and social media sites to social security data and police records to give every citizen a 'score' (linked by facial recognition). Depending on their score, citizens have more of less chance of securing a government job, a kindergarten place for their child, or a free annual medical check-up [28][29].

## 2.4 Cyberbody of cyberculture versus the virtual body of computer modelization

In contemporary art there are increasing opportunities to encounter cyber / cyborg bodies such those created by Mariko Mori, Yves Netzhammer, Tony Oursler, Zak Blas, John Rafman, Ed Atkins and Skawennati (to name but a few), but I would argue, still few digitized/virtual bodies. As explained by Mark Hansen in *Bodies in Code* [30], there is "an opposition between the "cyberbody of cyberculture" and the "virtual body of computer modelization."

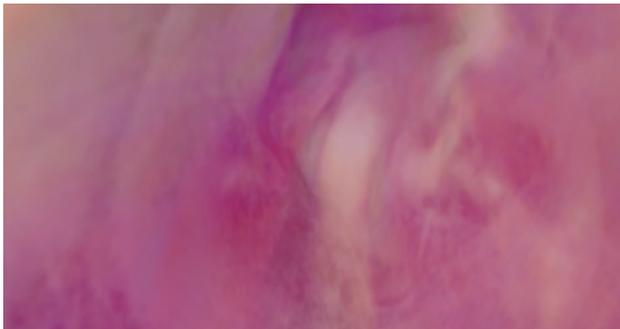

Figure 8: screen capture of *Deep Connection* showing the inside of the data body

MR scanners calculate water in the body by exciting and measuring the rotation of protons in hydrogen molecules. As human skin and hair is dry it does not scan. Cyber bodies are typically surfaced models – they have no interior and are all skin, whereas MR scans are all interior, with no skin. Although surface models of bodies are employed by many contemporary artists to ask important questions about digital culture, they do not in themselves address the technical, translational process of becoming digital, nor do they interrogate what it means to translate scan and identity data so that it can be 'expressed' generatively and interactively in an artistic virtual space. In fact, the limits of the MR scanned body are challenging to define, especially if there are breathing or cardiac artefacts, which raises exciting aesthetic questions about how to express the digital body into a 'virtual body' and visualise its boundaries and potential to change state. Karen Barad writes that "measurements are intra-actions (not interactions): the agencies of observation are inseparable from that which is observed. Measurements are world-making: matter and meaning do not pre-exist, but rather are co-constituted via measurement intra-actions" [31]. The intra-action of the body and the MR scanner proposes a world where the virtual body is akin to cloud with the potential to evaporate or condense, leak or float away. Where human are clouds of data that are constantly uploaded to or downloaded from. Illusionary harmless, seductive, amorphous clouds [32] that are in fact data corpuses. Bodies of data that can be dissected, reformatted, aggerated, duplicated and cross referenced so that they can be read and processed and machines [9][33] and ultimately, as Yuval Noah Harari warns, surveilled *under-the-skin* [34].

### 3.1 MAKING DEEP CONNECTION

*Deep Connection* was a collaborative art & science project made possible by bringing together the expertise of radiologists, biomedical researchers, computer scientists and artists. As the primary author this paper and subject of the MR scans, the dataset is related to as 'Oliver's', but the creation of *Deep Connection* a team project, hence the use of 'we' at other times when describing the creation of the work.

An increasing number of VR apps, such as Medical Holodeck (medicalholodeck.com), 3D Slicer (slicer.org), and the Body VR (thebodyvr.com), allow medical scans to be imported into virtual space. A review of the use of VR in medical data, however, shows that typically only sections of bodies are loaded into VR; bodies are fragmented and severed from their subject through anonymization protocols. Furthermore, the interpretation of these images is shaped by diagnostic language and sterile environments, which hinders more affective, social, and political readings of the body. For *Deep Connection* therefore, the first task was to acquire a high-resolution full body MR scan dataset. Furthermore, after having held a beating heart in my virtual hands in the *BodyVR* app and with the desire 'animate' a dataset, we also acquired 4D data of the heart beating so that it could be embedded perfectly into the 3D dataset.

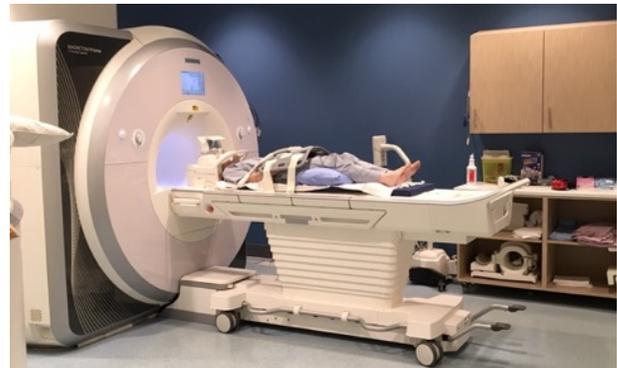

Figure 9: MR scanning at Peter S Allen MR Research Centre for *Deep Connection*

### 3.2 Scanning

In collaboration with Professor Richard Thompson and Peter Seres at the Peter S Allen MR Research Centre, University of Alberta, we worked to acquire sections of Oliver's body that later would be concatenated in the computing platform MATLAB. Unlike CT, which is able to scan the whole body (or most of the body depending on scanner bed movement and size of the body), MR scans are acquired in 30cm (or so) blocks. Due to the radiation emitted during a CT scan for they are not typically permitted for research. MR scanners essentially use super magnets to force protons in the body to align with a magnetic field. Then a radiofrequency is pulsed through the body of the patient so that the protons spin out of alignment of the magnetic field. The radiofrequency is then turned off and the protons in the body 'relax' to the original magnetic field. Measurements are taken at various points after the radiofrequency has been turned off [35]. For *Deep Connection*, images were acquired on a Siemens Prisma 3T MRI scanner (Erlangen, Germany) using three-dimensional (3D) T1-weighted protocols. T1 images

(measurements taken at timepoint 1) where chosen as they are the best for showing anatomical structures and soft tissue which gives the most 'realistic' rendering in VR. Scans were acquired at 1mm thick slices with 0.5mm in-plane resolution, the full body and real-time dynamic imaging (4D – x,y,z,t) was acquired for visualization of respiration and the beating heart.

Although these protocols gave the desired results, we did have to repeat the scans three times. During the first scan, the alarm buzzer that Oliver was holding imaged, making the dataset unusable. The second time, being too eager to outstretch her hand, it went out of the field of view of the scanner. Finally, on the third attempt, human error was minimized and the hand was imaged.

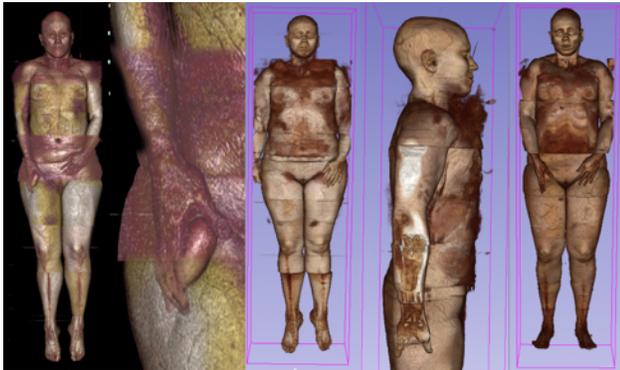

Figure 10: Renderings of the three concatenated datasets. Left is the first dataset were the alarm button was scanned, in the centre is the dataset where the hand was outside the scanning area and on the right is the final dataset.

### 3.3 Processing the data

For the next stage of the project Professor Kumar Puntihakumar and computer science graduate internship students Madhavi Nimalaratne and Preet Giri processed the data using ImageJ/Fiji, MATLAB and 3D Slicer. There was considerable overlap of slices between the blocks of data that had to be inspected and identified in Fiji. Once they were identified the overlapping slices were then deleted. Next, the slices were concatenated using MATLAB. One of the primary functions during the concatenation is fixing any misalignments in the scans. The full-body scan was performed in sections and the organs were shifted slightly in a few scans due to involuntary movements during scanning. We relied on the appearance of the structures in 2D slices to correct the misalignments by applying translational transformations manually. The 4D data was more complex to work with as all the slices had to sync together to a single heartbeat. The dynamic scans resulted in a series of folders of scans at different depths in the body but as each folder of images is acquired sequentially, the 'beat' and height of the breath occurred on different scans within the folders. We manually went through the scans, identifying the highest point of the breath (which became the first image in the folder) and lowest point of the breath (which became the last image in the folder), ensured there were the same number of scans in each folder (deleting and duplicating as necessary), and then concatenated the scans into one dataset using MATLAB.

These concatenated files were then imported into Slicer 3D where they were then cleaned of breathing artefacts and disturbing scan noise using the segmentation tools. Much to our delight and relief, this version of 3D Slicer, which was launched a few weeks after we started the project, had a VR plugin that allowed us to do much of the sketching and planning for the project in 3D Slicer itself. At present however, the 3D Slicer plugin does not allow for any interaction other than picking up the data and moving it so the final *Deep Connection* project had to be remade in the gaming engine Unity. The data was exported from the original DICOM format to MHD and RAW files using 3D Slicer and imported into Unity using a custom data reader code. We also performed a data conversion from the original single channel 16-bit to 8-bit format using a code written in Python programming language using the VTK module to allow for higher framerate rendering. The rendering of the datasets were performed in Unity using a custom raycasting shader code written by Professor Punithakumar. The rest of the functionalities related to the volume rendering such as the assignment of colour and opacity transfer functions were implemented in C# programming language.

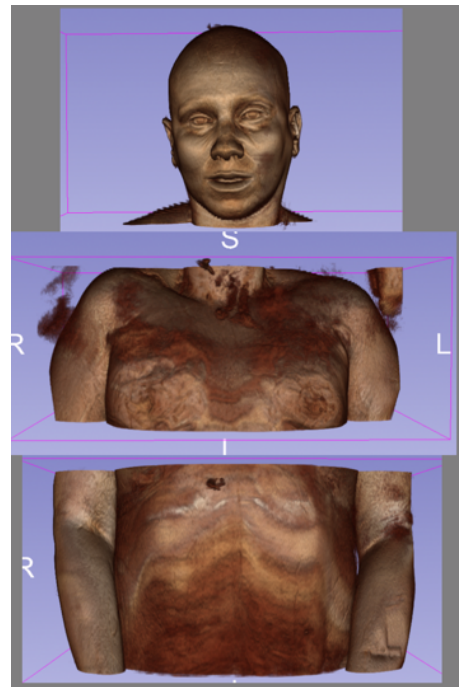

Figure 11: rendered data show three blocks that needed to be concatenated, equalized and cleaned of noise

### 3.4 Building the project in Unity

The 3D slicer VR plugin has an automatic clipping plane attached to the headset which means the data cuts away from the user as they put their head into it, recalling cross sectional medical scans. In Unity however, there is no automatic clipping plane so entering the data in Unity has a very different aesthetic. It looks (and feels) much like a galactic cloud, enhancing the feeling of entering into the space of the body, de-medicalising the data. In Unity interactions were coded to trigger the 4D dataset when the controller entered a collider placed on hand of the figure which then stopped when the controller exits the collider. The construction of the scene is very simple, there is no operating theatre, space ship or stone prison cell as is chillingly is the case in the VR app *Visible Human Dissector*. In *Deep Connection,* the data body floats in the dark void of an empty 0,0,0 value voxel [36].

The *Visible Human Dissector* VR app is an important one in this field both aesthetically, ethically and historically as it uses posthumous MR and CT scans of Joseph Paul Jernigan, the death row convict identified in the 1980s as the perfect specimen to become the 'Visible Human', the first digital anatomical atlas of a human created by the National Library of Medicine [8]. The first ever sculpture I made using medical scan data (called *I Know You Inside Out*) was with cryosections of Joseph Paul Jernigan and we are also using these scans in a related project which is explained in the future work section.

### 3.5 Sound

*Deep Connection* has an interactive soundscape composed and mixed by collaborator and sound artist Gary James Joynes. Three soundscape "stages" were created by combining recordings of the actual MR scanner in situ, modular synthesis composition and from my voice. Stage One consists of an imagined 'room' soundscape designed to hold sonic space when the user first puts on the headset and is outside the body. This soundscape beeps and whooshes utilizing the drone from the MR machine and its various scanning tones mixed with accompaniment by modular synthesizer programming - it is both mechanical, meditative and soporific. It signals to the user that they are in a space facilitated by mechanical processes. Once the user puts their head inside the body, the soundscape shifts as if filtered by flesh and body fluids with the pulsing of an imagined human heartbeat. The shift is reminiscent of putting your head underwater which also echoes how the data is in fact a representation of water. Finally, in Stage Three, as the user holds the figures hand a human voice sings a mourning song which lasts eight minutes. The mourning song is combination of lamenting cries inspired by the opening scene of *Blood Wedding* by Gabriel Garcia Lorca and parts of Faure's requiem. The recordings of the voice are highly emotionally charged, heightening the sense of the human emotion and connection as long as the human hold the virtual hand and 'animates' the scanned the body.

### 3.6 The sculptural installation

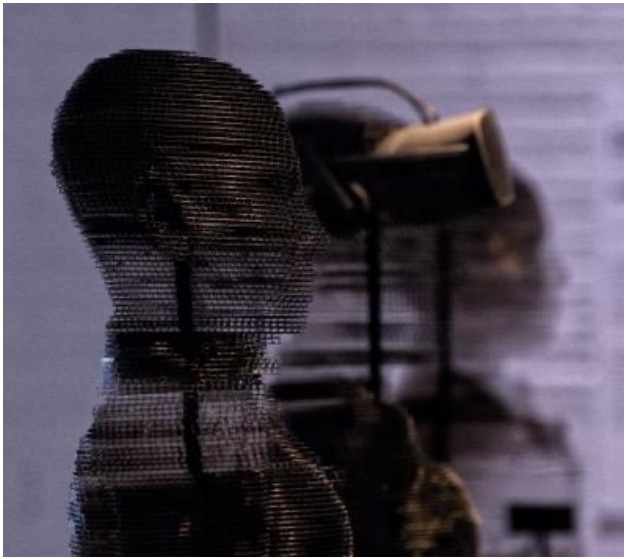

Figure 12: right view of Deep Connection sculptures showing how it is transparent/empty from the side.

Although at the time the failed scanning attempts were disappointing, in the long term the 'faulty' datasets afforded the possibility of making an installation of sculptures made from each of the datasets. The first two 'faulty' datasets have the Oculus Rift sensors in the chest (hardware which is no longer available, only its upgraded sensorless sibling the Rift S), and the final 'correct' dataset is the guardian of the workstation and holder of the headset. The sculptures are made from layers of laser cut coroplast. Paths for the laser cutting were generated by sectioning 3D surface models rendered from the MR scan data. Black coroplast was chosen for its colour and plasticity so that it would camouflage the hardware and also for being opaque when viewed on the front edge and transparent from the side.

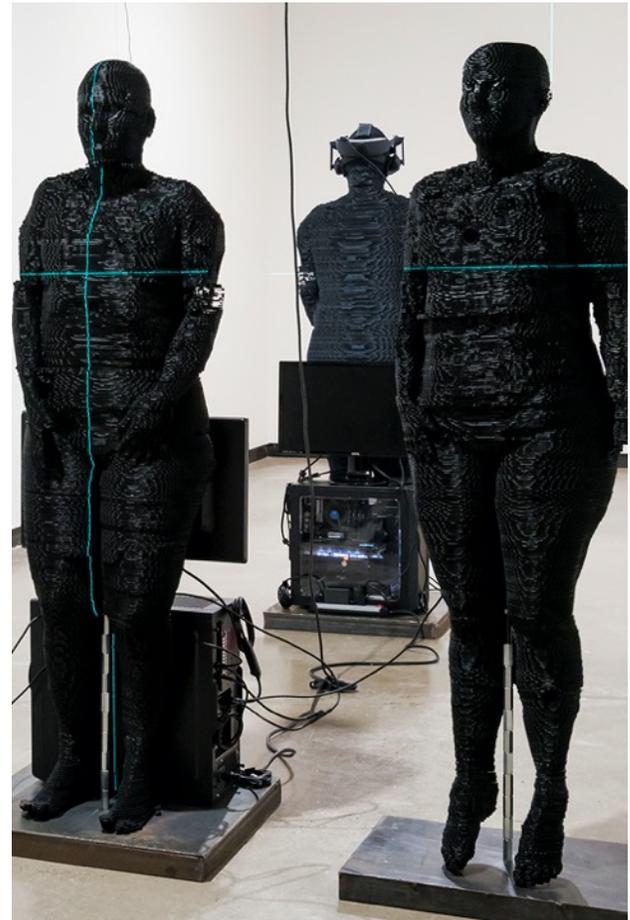

Figure 13: Deep Connection sculptures. Left sculpture 'guards' the workstation and the left sculpture has the sensor embedded in its chest.

In creating the sculptural installation, I was interested initially in how the faulty data enabled the final dataset and in drawing attention to how much data is discarded if it is not deemed 'correct'. This developed into a studio-based exploration of human relationship (specifically the artist/craftsman) with digital objects, as well as the objectification of data and the datafication of objects as explored by Hui Yuk in *On the Existence of Digital Objects* [37]. In his book Yuk puts forward that 'digital objects' (such as data and meta data) are different from 'technical objects' and 'the double movement from object to data, from data to object will be an ongoing project'. Later he writes ' a digital object is also constantly in the process of re-establishing and renegotiating its relations with other objects, systems, and users within their associated milieux.

When the Deep Connection installation was first exhibited in 2019, Oliver was present in the gallery as an 'assistant' to visitors (in that she explained the app to them before they put on the headset and watched them to make sure they didn't trip or bump into the sculptures) and the technology (in that she turned it on, pressed the necessary buttons, clicked the right icons to launch the VR app). This created a scenario where the original scanned object (Oliver's body) was present alongside the digital object generated by her body (the scan dataset), the physical life size sculptures generated by the scan dataset (in collaboration with the artist's body and digitally mechanised machine) [38][39], and finally, the body of the viewer who generates the visualisation as they move the headset and the controllers. This complex, intimate entanglement and interdependence of humans, digital objects and technology, is one that as Yuk warns, will be with us to consider, question and negotiate for several decades.

**4 CONCLUSION AND FUTURE WORK**

With VR, the viewer's physical body becomes the interface with the artwork. Rather than encourage disembodiment, as is often the fear with new technologies, VR elevates and re-positions embodiment, compelling us to reconsider fundamental phenomenological, perceptual, spatial, and aesthetic issues [30][40]. In *Deep Connection*, viewers had to lean down and put their heads into 'Olivers' chest before shuffling bent over through the rest of 'Oliver's' body. In the public space of the exhibition, their movements became part of the installation; their intrusions into a female body, made possible through a series of complex phallogocentric technological processes, were rendered visible and spectacular. Others watched them move uncannily through a virtual space that they could only imagine, until it was their turn to put on the VR headset and participate. Public exhibition of *Deep Connection* highlighted a novel and fascinating set of aesthetic considerations about viewing medical data in VR, which move beyond concerns with the appearance of the scanned body to include the rituals of the body as interface with, and intruder into, data.

To lay oneself open, to become so transparent and open to scrutiny is potentially terrifying and dangerous, especially in a world where we are being increasingly indoctrinated to criticize and police each other, and when large corporations and governments motivation's for collecting data are driven primarily by capitalistic, surveilling and controlling motives [41][42]. As we now plan new artworks with medical data and VR, we need to ask whether the rules for what we do and make transparent/public with our own data (if indeed it is ours), are different to what we do to other people's data (if indeed it is theirs'?) The general rule when working scientifically with scan data is that if it is 'anonymized' or 'de-identified', it can be freely used for research [43]. Traditionally scan anonymization involved removing all textual personal identifiers, but now that high resolution scans can be rendered to create recognizable likeness to the scan subject, scientific anonymization protocols now involve 'de-facing' data [44]. Is this also the best solution ethically and aesthetically when making artworks? What should users/viewers be allowed to do with and in the data of others?

The *Deep Connection* project has led to the establishment of a larger interdisciplinary team project called 'Know Thyself as a Virtual Reality' that now includes researchers from law, art history, philosophy and digital humanities, rehabilitation medicine and kinesiology as well as arts, radiology and computer science (https://www.knowthyself.ualberta.ca/). Together we are working to developing both a set of open source tools that will allow other artists and medical students to work creatively with scan data in VR, as well as a set of ethical guidelines for their use. This project takes the form of interdisciplinary presentations and symposiums as well as continuing to developing VR artworks made with medical scan data. The two VR projects we are currently developing are called *My Data Body* and *Your Data Body*. *My Data Body* has at its centre the same MR scan dataset used in *Deep Connection*. Embedded into the semi-transparent, virtual body are other data corpuses downloaded from Facebook and Google. These textual data corpuses are plotted into cross sections of the body. In the horizontal (axial) plane, Mac terminal data is plotted into bone, Google data into muscle and Facebook data into fat. In the vertical (coronal) plane, data usage agreements are plotted and into the depth (sagittal) plane, are theoretical texts about virtuality and privacy in the digital age. The viewer can pull out these cross sections and read them; once they let go the cross sections float away but ultimately and uncontrollably return to the scanned body. Passwords and logins flow back and forth through veins and arteries and hashtags pool in organs. Certain organs can be pulled out of the body and 'drawn with': the heart leaves a trail of emojis and the brain a trail of login pop-up windows demanding usernames and passwords. The medically scanned, passive/obedient semi-transparent body becomes a data processing site that can be pulled apart and dis-'organised'. The whole body/data processing site finds itself at the centre of a data cloud generated from social media data.

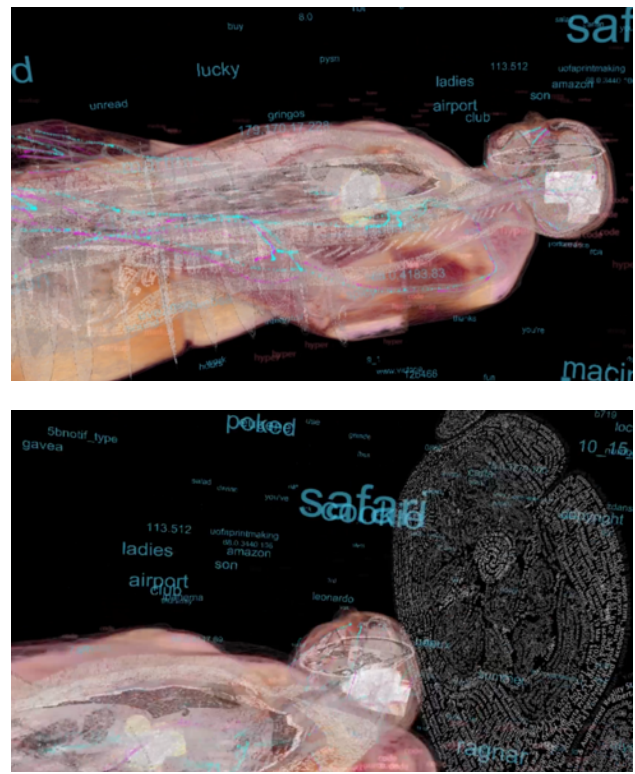

Figure 14&15: screen capture of My Data Body VR project showing cross sections of the body made of Facebook, Google and Mac terminal data

*Your Data Body* is a partner project to *My Data Body* made using a combination of open source and donated datasets. This project focuses on issues of data privacy and ownership, playing on the etymology of the word data meaning 'given'. The user can pick up and move datasets around. The scanned body parts can resized and re-colored, inviting a playful stacking of the body parts to make a whole Frankenstein-like figure. Attached to each body part is an audio file which is triggered when the user holds and manipulates it. Anonymized open source datasets are accompanied by an automated voice which reads the study data published alongside the dataset, whereas the donated datasets will have different conversational AI characters/chatbot with whom user can 'discuss' different issues relating to data ownership, privacy and virtuality with.

ACKNOWLEDGEMENTS

Special thanks to Dr Richard Thompson (University of Alberta) and Dr Chris Hanstock (University of Alberta) for their guidance and support in the acquisition of the MR scan data, Professor Pierre Boulanger of the Advanced Man and Machine Interface (AMMI) lab, (University of Alberta) for their continued support and encouragement and Katrina Ingram of Ethically Aligned AI for her expertise on health data ethics and artificial intelligence. Sincere thanks also to the University of Alberta Computer Science Multimedia internship students Madhavi Nilmarte and Preet Giri who worked on the *Deep Connection* project.

I would also like to acknowledge the support from Peter S Allen MR Scanning Centre, President's Grants for the Creative and Performing Arts from the Killam Research Fund and Kule Institute for Advanced Studies, University of Alberta, Canada.

REFERENCES

[1] M. Heidegger. *The Question Concerning Technology and Other Essays*. Harper and Row, 1977.
[2] Thamil Amudhu, L. B. "A Review on the Use of Socially Assistive Robots in Education and Elderly Care." Materials Today: Proceedings, Jan. 2020.
[3] S. Turkle. Robot Companions in 'What Scientific Idea is Ready for Retirement?, Edge.org 2016. https://www.edge.org/response-detail/25420 [accessed 28th May 2021]
[4] L. Billingsley, Using Video Conferencing Applications to Share the Death Experience During the COVID-19 Pandemic. In *Journal of Radiology Nursing*, volume 39, issue 4, pages 275-277, 2020.
[5] S. Zhang. The Pandemic Broke End-of-Life Care in *The Atlantic*, June 2020.
[6] J. Sawday. *The Body Emblazoned: Dissection and the Human Body in Renaissance Culture*. Routledge, 1995.
[7] J. Van Dijk. *The Transparent Body: A Cultural Analysis of Medical Imaging (In Vivo)*, University of Washington Press, 2006
[8] C. Waldby. *The Visible Human Project: Informatic Bodies and Posthuman Medicine*. Routledge 2000.
[9] A. Cunningham. The End of the Sacred Ritual of Anatomy in *Canadian Bulletin of Medical History*, volume 18, no 2, Fall 2001, pages 187-204, 2001.
[10] R, Cuir. *"Know thyself"*, Art and Anatomy: A Discontinued History. 7th February, Courtauld Institute. London, 2008.
[11] D. Lupton. *The Quantified Self*. Malden, MA: Polity, 2016.
[12] S. Casini. *Giving Bodies Back to Data: Image Makers, Bricolage, and Reinvention in Magnetic Resonance Technology*, MIT Press. 2021.
[13] M. Kemp. Science in culture. In *Nature* 424, 18, 2003
[14] J. Prophet, (Projection) mapping the brain: a critical cartographic approach to the artist's use of fMRI to study the contemplation of death. Art Paper, SIGGRAPH Asia, Kobe, Japan, 2015
[15] Catherine Monahon and Elizabeth Jameson. Intimate Visions: Representations of the Imperfect Body in the Age of Digital Medicine in *Leonardo* 2020; 53 (3): 281–287
[16] D.Stahl. Imaging and Imagining Illness: Becoming Whole in a Broken Body, Wipf & Stock, 2018.
[17] H. Moravec. Mind Children: the Future of Robot and Human Intelligence. Harvard University Press 1988.
[18] Medical Futurist. (2018, August 17). Google's Masterplan in Healthcare. Retrieved from - https://www.youtube.com/watch?v=VsLik7rQY6A
[19] F. Densford. (2018, September 20). Report: Google's Nest looking to expand into healthcare. In *Mass Device*. Retrieved from - https://www.massdevice.com/report-googles-nest-looking-to-expand-into-healthcare/
[20] D. Phelan. (2019, November 1). Google buys Fitbit for $2.1 billion: Here's what it means. Forbes. Retrieved from - https://www.forbes.com/sites/davidphelan/2019/11/01/google-buys-fitbit-for-21-billion-heres-what-it-means/?sh=739faa02732f
[21] L. Lovett. (2021, February 23). Google rolls out EHR navigation tool Care Studio. Mobi Health News. Retrieved from - https://www.mobihealthnews.com/news/google-rolls-out-ehr-navigation-tool-care-studio
[22] B. Latour. *Pandora's Hope: Essays On the Reality of Science Studies*. Harvard University Press, 1999.
[23] S. Casini. From Where Do We See? Opening Up The Black Box of Biomedical Imaging in *Immobile Choreography* (pp.31-66), Grampian Hospitals Art Trust. 2019.
[24] J. Slatman. Transparent Bodies: Revealing the Myth of Interiority, in R. van de Vall & R. Zwijnenberg (eds.). *The Body Within: Art, Medicine and Visualization*, Brill, Leiden, 2009 pp. 107-122
[25] L. Cartwright. *Screening the Body: Tracing Medicine's Visual Culture*, University of Minnesota Press, 1995.
[26] L. Scott. *The Four-Dimensional Human: Ways of Being in the Digital World*. William Heinemann, 2015.
[27] D. Eggers. *The Circle*, Penguin, 2014.
[28] C. Campbell. How China Is Using "Social Credit Scores" to Reward and Punish Its Citizens, in *Time*, Davos 2019
[29] P. Poornima. "Confusion clouds China's social credit system", *Japan Times*, Sept 3, 2019
[30] M. Hansen. Bodies in Code: Interfaces with Digital Media. Routledge 2006.
[31] K. Barad, *Meeting the Universe Halfway: Quantum Physics and the Entanglement of Matter and Meaning*. Duke University Press, 2007
[32] T. Hu. *A Prehistory of the Cloud*, MIT Press, 2015
[33] Baser et al. *The Age of You*, Museum of Contemporary Art Toronto, 2019
[34] Y. Harari. The World After Coronavirus in *The Financial Times*. March 19 2020.
[35] A. Berger. "Magnetic resonance imaging." *BMJ (Clinical research ed.)* vol. 324,7328 (2002): 35. doi:10.1136/bmj.324.7328.35
[36] S. Cubitt. 2014. The Practice Of Light: A Genealogy Of Visual Technologies From Prints To Pixels, The MIT Press.
[37] H. Yuk. *On the Existence of Digital Objects*, Minnesota Press, 2016.
[38] K. Hayles. How We Became Posthuman: Virtual Bodies in Cybernetics, Literature and Informatics, University of Chicago Press 1999.
[39] G. Simondon, *Du mode d'existence des objets techniques*. Aubier, 2012.
[40] R. Diodato. *Aesthetics of the Virtual*, State University of New York Press. 2012.
[41] R. Benjamin. *Captivating Technology, Race, Carceral Technoscience, and Liberatory Imagination in Everyday Life*. Duke University Press. 2019.
[42] C. O'Neil. *Weapons of Math Destruction: How Big Data increases Inequality and Threatens Democracy*. Crown. 2016.
[43] Chevrier et al Use and Understanding of Anonymization and De-Identification in the Biomedical literature: Scoping Review in *J Med Internet Res*. Vol. 21, 5, May 2019.
[44] J. Muschelli. Recommendations for Processing Head CT Data in *Front Neuroinform*. Vol 13, no 61. 2019.